\documentclass[letterpaper,peerreview,final,12pt]{IEEEtran}
\usepackage[T1]{fontenc}
\usepackage{type1cm}
\usepackage[cmex10]{amsmath}
\usepackage{amsthm} \theoremstyle{definition}
\usepackage[varg]{txfonts}
\usepackage[normalem]{ulem}
\usepackage[]{cite}
\usepackage{multirow,bigdelim}
\usepackage{url}
\usepackage{color}

\newcommand{\rk}{\textsf{rk}}

\theoremstyle{plain}
\newtheorem{thm}{Theorem}[section]
\newtheorem*{theo}{Theorem}

\newtheorem{defn}{Definition}
\newtheorem{lem}[thm]{Lemma}
\newtheorem{prop}[thm]{Proposition}
\newtheorem{cor}[thm]{Corollary}

\theoremstyle{definition}

\newtheorem*{dem}{Proof}

\theoremstyle{remark}

\begin{document}
\title{Generalized rank weights: a duality statement}
\author{J\'er\^ome~Ducoat\\
\vspace{.2cm}
\small{Division of Mathematical Sciences,\\ School of Physical and Mathematical Sciences,\\ Nanyang Technological University, Singapore\\
\textit{Email: jducoat@ntu.edu.sg}\\
\vspace{.5cm}
}
}
\maketitle

\begin{abstract} We consider linear codes over some fixed finite field extension $\mathbb F_{q^m}/\mathbb F_q$, where $\mathbb F_q$ is an arbitrary  finite field. In \cite{gabi}, Gabidulin introduced rank metric codes, by endowing linear codes over $\mathbb F_{q^m}$ with a rank weight over $\mathbb F_q$ and studied their basic properties in analogy with linear codes and the classical Hamming distance. Inspired by the characterization of the security in wiretap II codes in terms of generalized Hamming weights by Wei \cite{wei}, Kurihara \textit{et al.} defined in \cite{kuri} some generalized rank weights and showed their relevance for secure network coding. In this paper, we derive a statement for generalized rank weights of the dual code, completely analogous to Wei's one for generalized Hamming weights and we characterize the equality case of the $r^{th}$-generalized Singleton bound for the generalized rank weights, in terms of the rank weight of the dual code.
\end{abstract}

\section{Introduction}
Let $q$ be the power of some prime number, let $m\geq 1$. We denote by $\mathbb F_q$ (resp. $\mathbb F_{q^m}$) the field (unique up to isomorphism) with $q$ (resp. $q^m$) elements. Then $\mathbb F_{q^m}/\mathbb F_{q}$ is a field extension of degree $m$. \\

Let $n\geq 1$ and consider the vector space $\mathbb F_{q^m}^n$. Let $(u_1,....,u_m)$ be a basis of $\mathbb F_{q^m}$, seen as an $m$-dimensional vector space over $\mathbb F_q$. For every $x=[x_1,...,x_n]\in \mathbb F_{q^m}^n$, there exist some coefficients $x_{j,i}\in \mathbb F_q$ for $1\leq i\leq n$ and $1\leq j\leq m$ such that, for every $1\leq i \leq n$, $$x_i=\overset{m}{\underset{j=1}{\sum}} x_{j,i}u_j.$$ We then set $$\lambda(x)=\begin{bmatrix}x_{i,j}\end{bmatrix}\in \mathcal M\textrm{at}_{n,m}(\mathbb F_q).$$

Let $C$ be a linear code over $\mathbb F_{q^m}$ of length $n$ (i.e. a vector subspace of $\mathbb F_{q^m}^n$). Gabidulin (\hspace{-.1mm}\cite{gabi} and also Roth independently, \cite{roth}) defined the rank of a vector $x\in \mathbb F_{q^m}^n$  (denoted by $\rk(x)$) to be the rank of $\lambda(x)$, the rank distance between two codewords $x,y\in C$ to be $\rk(x-y)$ and the rank weight of $C$ by $$d(\lambda(C))=\underset{x\in C\setminus \{0\}}{\min} \rk(x).$$ 

In \cite{silv}, in the framework of linear network coding, Silva and Kschischang proposed the rank distance to characterize when wiretap network codes achieve perfect secrecy.\\ 

A natural question arose then, about the existence of generalized rank weights, in analogy with the generalized Hamming weights defined by Wei in \cite{wei}, known to describe the equivocation of the eavesdropper for wiretap II codes.\\

A first step in this direction was given by Oggier and Sboui \cite{fred} and was completed independently in \cite{kuri}, by Kurihara, Matsumoto and Uyematsu. We first introduce some tools. For every $x=[x_1,...,x_n]\in \mathbb F_{q^m}^n$, we denote by $x^q$ the vector $[x_1^q,...,x_n^q]$. For every vector subspace $V\subset \mathbb F_{q^m}^n$, we set $V^q=\{x^q \mid x\in V\}$.\\

We then consider the set $\Gamma(\mathbb F_{q^m}^n)=\{V\subset \mathbb F_{q^m}^n \mid V^q=V\}$. For every vector subspace $V$ of $\mathbb F_{q^m}^n$, we set $$V^*= \overset{m-1}{\underset{j=0}{\sum}} V^{q^j}.$$ Then $V^*$ is the smallest subspace containing $V$ and belonging to $\Gamma(\mathbb F_{q^m}^n)$.\\

Recall that $C$ is a linear code over $\mathbb F_{q^m}$ of length $n$. Let $k$ be its dimension. For every $1\leq r\leq k$, a refinement of the definition proposed by Oggier and Sboui for the $r^{th}$-generalized rank weight in \cite{fred} is  $$ d_r\left(\lambda(C)\right)=\underset{\dim D=r}{\underset{D\subset C}{\min}} \underset{x\in D^*}{\max} \hspace{.2cm} \rk\left( \lambda(x)\right)$$
and the definition proposed by Kurihara, Matsumoto and Uyematsu in \cite{kuri} is   
$$\mathcal M_r(C)=\underset{\dim(C\cap V)\geq r}{\underset{V\in \Gamma(\mathbb F_{q^m}^n)}{\min}} \dim V .$$

Notice that the $D^*$ involved in the first definition means the smallest subspace containing $D$ and stable by the $q$-power componentwise, as defined above.\\

We let the reader note that these two definitions are given in analogy with the $r^{th}$-generalized Hamming weight, defined as follows by Wei in \cite{wei} : for every $1\leq r \leq k$, $$d_r(C)=\underset{\dim D=r}{\underset{D\subset C}{\min}}\underset{x\in D}{\max} \mid \textsf{Supp}(D)\mid  = \underset{\dim(V\cap C)\geq r}{\underset{V\in \Lambda(\mathbb F_{q^m}^n)}{\min}} \dim V,$$

where $\textsf{Supp}(D)=\{i\in\{1,...,n\} \mid \exists x=[x_1,...,x_n]\in D, x_i\neq 0\}$, $\mid . \mid$ denotes the order of a set, and $\Lambda(\mathbb F_{q^m}^n)$ is the set of the vector subspaces of $\mathbb F_{q^m}^n$, generated by elements of the canonical basis. Note that the right equality is easy to check in that case.\\

Kurihara, Matsumoto and Uyematsu proved the following (\hspace{-0.1mm}\cite{kuri}, Lemma 11).
\begin{prop}\mbox{\null}
\label{d1=m1}
\begin{center}
Let $n\leq m$. For every $x\in \mathbb F_{q^m}^n$, $\dim\left(\langle x \rangle^*\right)=\rk\left(\lambda(x)\right) .$ \end{center}
\end{prop}

\vspace{1cm}
This immediately shows that $ \mathcal M_1(C)=d(\lambda(C))=d_1(\lambda(C)).$ In Section II, we prove that $ \mathcal M_r(C)=d_r(\lambda(C))$ for every $1\leq r\leq k$ in the case where $n\leq m$. \\ 

In \cite{kuri}, Kurihara, Matsumoto and Uyematsu proved the following monotonicity property (\hspace{-0.1mm}\cite{kuri}, Lemma 9):
\begin{thm}
\label{t4}
We have $1\leq \mathcal M_1(C)< \mathcal M_2(C) <...< \mathcal M_k(C)\leq n$.
\end{thm}

We also give in Section II a different proof of this statement. Note that the monotonicity property legitimates these two definitions as a suitable candidate for the notion of generalized rank weight. \\

In Section III, we continue the analogy with generalized Hamming weights, extending to generalized rank weights the statement that Wei proved in \cite{wei}, Theorem 3. Let $C^\perp$ denote the dual code, that is to say the orthogonal vector subspace with respect to the usual bilinear form $$\langle.,.\rangle:([x_1,...,x_n],[y_1,...,y_n])\mapsto \overset{n}{\underset{i=1}{\sum}}x_iy_i.$$ We then link the generalized rank weights of the dual code $C^\perp$ to the generalized rank weights of $C$:
\begin{thm}
\label{t1}
Let $C$ be a linear code of dimension $k$ over $\mathbb F_{q^m}$ and of length $n$. Then $$\{\mathcal M_r(C)\mid 1\leq r\leq k\}= \{1,...,n\}\setminus \{n+1-\mathcal M_r(C^\perp) \mid 1\leq r \leq n-k\}.$$ 
\end{thm}

As a consequence of this statement, we end this paper by deriving a characterization of the equality case in the $r^{th}$-generalized Singleton bound for the generalized rank weights (\hspace{-.1mm}\cite{kuri}, Proposition 10), in terms of the rank weight of the dual code.

%
%


\section{General properties for the generalized rank weights}
The aim of this section is to prove that both previously proposed generalized weights are the same.
\begin{prop}
\label{p1}
Let $n\leq m$. For every $1\leq r \leq k$, $d_r\left(\lambda(C)\right)= \mathcal M_r(C).$\\
\end{prop}

\begin{dem}
Let us first prove that $d_r\left(\lambda(C)\right)\leq \mathcal M_r(C).$ Let $V\in \Gamma(\mathbb F_{q^m}^n)$ such that $\dim\left(C\cap V\right)\geq r$. Let $D$ be a subspace of $C\cap V$ of dimension $r$. For every $x\in D^*$, by Proposition \ref{d1=m1}, $$\dim\left(\langle x \rangle^*\right)=\rk\left(\lambda(x)\right).$$ 

Since $D^*$ is the smallest invariant subset containing $D$, then $D^*\subset V$, so $x\in V$ and since $V$ is invariant by the elevation to the power $q$, we have $\langle x \rangle^*\subset V$, so $\dim\left(\langle x \rangle^*\right)\leq \dim V$. Hence, for every $x\in D^*$, $\rk\left(\lambda(x)\right)\leq \dim V$, thus $$\underset{x\in D^*}{\max} \hspace{.2cm} \rk\left(\lambda(x)\right) \leq \dim V.$$
Therefore, $$ d_r\left(\lambda(C)\right)\leq \dim V.$$

Since this inequality is true for every invariant subspace $V$ such that $\dim(V\cap C)\geq r$, we get that $$d_r\left(\lambda(C)\right)\leq \mathcal M_r(C).$$ 

We now come to the converse inequality. It follows from the following lemma :
\begin{lem}
\label{l1}
Assume that $n\leq m$. Let $V\in \Gamma(\mathbb F_{q^m}^n)$. Then there exists $x\in V$ such that $V=\langle x\rangle^*$.\\
\end{lem}

\begin{dem}
 Let $l$ be the dimension of $V$. Then there exists some basis $(e_1,...,e_l)$ of $V$ coming from $\mathbb F_q$ (i.e. every coefficient of the $e_i$ belongs to $\mathbb F_q$, see \cite{stich}, Lemma 1). Let $x\in V$ with coefficients $x_1,...,x_l$ when $x$ is decomposed in the basis $(e_1,...,e_l)$ (these coefficients belong to $\mathbb F_{q^m}$). Assume that the family $(x_1,...,x_l)$ is free over $\mathbb F_q$. Then a vector $y=\underset{i=1}{\overset{l}{\sum}}y_i e_i$ in $V$ belongs to $\langle x\rangle ^*$ if and only if there exist some $\mu_0,...,\mu_{m-1}\in \mathbb F_{q^m}$ such that, for every $i=1...l$, $$y_i=\underset{j=0}{\overset{m-1}{\sum}} \mu_{j}x_i^{q^j},$$ which is equivalent to $$\begin{bmatrix}y_1\\ \vdots \\y_l\end{bmatrix} = \begin{bmatrix} x_1 & x_1^q & \cdots & x_1^{q^{m-1}}\\ \vdots & \vdots &  & \vdots \\ x_l & x_l^q & \cdots & x_l^{q^{m-1}}\end{bmatrix} \begin{bmatrix} \mu_0 \\ \vdots \\ \mu_{m-1}\end{bmatrix}.$$
 
Since the family $(x_1,...,x_l)$ is free over $\mathbb F_q$, the matrix $$\begin{bmatrix} x_1 & x_1^q & \cdots & x_1^{q^{m-1}}\\ \vdots & \vdots &  & \vdots \\ x_l & x_l^q & \cdots & x_l^{q^{m-1}}\end{bmatrix}$$ has maximal rank $l$. Therefore, $\dim (\langle x^*\rangle)=l=\dim V$, which proves that $V=\langle x^*\rangle. \square$\\

\end{dem}

This completes the proof of Proposition \ref{p1}.
$\square$ \\
\end{dem}


We continue Section II by giving another proof of the monotonicity property, already stated by Kurihara, Matsumoto and Uyematsu (\cite{kuri}, Lemma 9). More precisely, we prove here the following proposition.
\begin{prop}
\label{l3} 
Let $C$ be a linear code of dimension $k$ and length $n$ over $\mathbb F_{q^m}$. Then, for every $1< r \leq k$, $$(q^{mr}-1)\mathcal M_{r-1}(C)\leq (q^{mr}-q^m)\mathcal M_r(C).$$
\end{prop}

\begin{dem}
Let $1< r \leq k$. Let $t$ denote the quotient $\frac{q^{mr}-1}{q^{m}-1}$. It is well-known that $t$ is the number of $(r-1)$-dimensional subspaces in a vector space of dimension $r$ over $\mathbb F_{q^m}$ (see for instance \cite{huff} Exercise 431). 

Let $D$ be an $r$-dimensional subspace of $C$ such that $\mathcal M_r(C)=\dim D^*$. We enumerate by $D_1,...,D_t$ the list of all the $(r-1)$-dimensional subspaces of $D$. \\

We want to show that $$(q^{mr}-1) \mathcal M_{r-1}(C) \leq (q^{mr}-q^m) \mathcal M_r(C),$$ i.e. that $$(q^{mr}-1)\left(\mathcal M_r(C)- \mathcal M_{r-1}(C)\right) \geq (q^{m}-1) \mathcal M_r(C),$$ which is equivalent to $$t \left( \mathcal M_r(C)- \mathcal M_{r-1}(C) \right) \geq \mathcal M_r(C).$$ Moreover, $\mathcal M_r(C)= \dim D^*$ and for every $1\leq i\leq t$, $\dim D_i^*\geq \mathcal M_{r-1}(C)$, so it is enough to prove that \begin{equation} \label{eq3} \underset{i=1}{\overset{t}{\sum}} (\dim D^* - \dim D_i^*)\geq \dim D^*. \end{equation}

Set $s=\dim D^*$. Since $D^*$ belongs to $\Gamma(\mathbb F_{q^m}^n)$, we can find a basis $(e_1,...,e_s)$ of elements which have coordinates in $\mathbb F_{q}$ (see \cite{stich}, Lemma 1). For $1\leq j\leq s$, let $V_j$ be the $(s-1)$-dimensional subspace of $D^*$ generated by the family $(e_1,...,\widehat{e_j},...,e_{s})$, where the $\widehat{e_j}$ means that the vector $e_j$ is excluded from this family. These vector spaces $V_j$ belong to $\Gamma(\mathbb F_{q^m}^n)$ (since they have a basis with coordinates in $\mathbb F_q$) and have dimension $s-1$. \\

Let $1\leq j\leq s$ and consider the intersection $V_j\cap D$. Then $V_j\cap D\subsetneq D$ (otherwise it would contradict the minimality of $\dim D^*$). Since $D\not\subset V_j$, $\dim(V_j+D)> \dim V_j$, then $\dim (V_j+D)=\dim D^*=s$ and we have $$\dim(V_j\cap D)=\dim V_j+\dim D-\dim (V_j+D)=s-1+\dim D-s=\dim D-1=r-1.$$

Therefore, there exists $i_j\in \{1,...,t\}$ such that $D_{i_j}=V_j\cap D$. Here we catch the reader's attention on the fact that the $i_j$ might be the same for different indices $j$. Up to reindexing the basis $(e_1,...,e_s)$ (and hence the subspaces $V_1,...,V_s$), we can assume that there exist some integers $t_1,...,t_s$ such that \begin{center} for every $1\leq l\leq t_1$, $V_l\cap D=D_{i_{t_1}}$,\\ \vspace{.1cm} for every $t_1+1\leq l \leq t_2$, $V_l\cap D=D_{i_{t_2}}$,\\ ... \\ for every $t_s+1\leq l \leq t_s=s$, $V_l\cap D=D_{i_{t_s}}$, \end{center} with the subspaces $D_{i_{t_1}},...,D_{i_{t_s}}$ two by two distinct. \\

Thus, we have, for every $1\leq j\leq s$, $D_{i_{t_j}}^*\subset V_{t_{j-1}+1}\cap \cdots\cap V_{t_j}$ (with the convention that $t_0=0$) and taking dimensions, $$\dim D_{i_{t_j}}^* \leq s-(t_j-t_{j-1}).$$

Therefore, $$\underset{i=1}{\overset{s}{\sum}} (\dim D^* - \dim D_{i_{t_j}}^*) \geq \underset{i=1}{\overset{s}{\sum}} (t_j-t_{j-1}) = t_s-t_0=s.$$ Since we have the obvious inequality $$\underset{i=1}{\overset{s}{\sum}} (\dim D^* - \dim D_{i_{t_j}}^*)\leq \underset{i=1}{\overset{t}{\sum}} (\dim D^* - \dim D_i^*),$$ Inequality (\ref{eq3}) holds, which completes the proof of Proposition \ref{l3}.  
 $\square$\\
\end{dem}

As an immediate consequence of the monotonicity property (Theorem \ref{t4}), Kurihara, Matsumoto and Uyematsu  stated that the generalized Singleton bounds hold for generalized rank weights (\hspace{-.1mm}\cite{kuri}, Proposition 10).
\begin{cor}
Keeping the notation above, let $1\leq r\leq k$. Then, we have $$\mathcal M_r(C) \leq n-k+r.$$ 
\end{cor}

We also remark here that it directly followed from the fact that for every $1\leq r\leq k$, $\mathcal M_r(C)$ is always lower than or equal to the $r^{th}$-generalized Hamming weight.\\

\begin{defn}
\label{rthrankMRD} Keeping the notation above, we say that a linear code $C$ of dimension $k$ and length $n$ over $\mathbb F_{q^m}$ is $r^{th}$-rank MRD (or in short $r$-MRD) if we have $\mathcal M_r(C)=n-k+r$. \\
\end{defn}

At the end of Section III, we give a characterization for a code to be $r$-MRD in terms of the (first) rank distance of its dual code $C^\perp$.\\

Note also that for (generalized) Hamming weights, a refinement of the (generalized) Singleton bound, called Griesmer bound holds (see for instance \cite{huff}, Theorem 7.10.10). It is then natural to wonder whether such analogous bounds hold for the generalized rank weights. The answer is positive but due to the constraints on $q$, $m$ and $n$, these bounds are exactly identical to the generalized Singleton bounds.\\

\section{Duality and generalized rank weights : proof of Theorem \ref{t1}}
Recall that the dual (orthogonal) code of $C$, denoted by $C^\perp$, is defined as $$C^\perp=\{x\in \mathbb F_{q^m}^n\mid \forall y\in C, \langle x,y\rangle=0\},$$ where $\langle.,.\rangle$ is the bilinear form defined in Section I. We state the following lemma :
\begin{lem}
\label{l5} Let $V\in \Gamma(\mathbb F_{q^m}^n)$. Then $V^\perp\in \Gamma(\mathbb F_{q^m}^n)$.
\end{lem}

\begin{dem}
Let $x\in V^\perp$. We need to show that $x^q\in V^\perp$. Then, let $y\in V$. Let us prove that $\langle x^q,y \rangle=0$. Since $y=[y_1,...,y_n]\in V=V^q$, there exists some $z=[z_1,...,z_n]\in V$ such that $y=z^q$. Hence, we have $$\begin{aligned} \underset{1\leq i\leq n}{\sum} x_i^qy_i &=  \underset{1\leq i\leq n}{\sum} x_i^qz_i^q \\ & = \left(\underset{1\leq i\leq n}{\sum} x_iz_i\right)^q\\ &=0^q =0, \end{aligned}$$  
which completes the proof. $\square$\\
\end{dem}

Let us recall the statement of Theorem \ref{t1}, which we are to prove here.
\begin{theo}
Let $C$ be a linear code of dimension $k$ over $\mathbb F_{q^m}$ and of length $n$. Then $$\{\mathcal M_r(C)\mid 1\leq r\leq k\}= \{1,...,n\}\setminus \{n+1-\mathcal M_r(C^\perp) \mid 1\leq r \leq n-k\}.$$ 
\end{theo}

\begin{dem}
We start with stating the following lemma :
\begin{lem} \label{l2} Let $1\leq r\leq n-k$ and let $t=k+r-\mathcal M_r(C^\perp)$. Then,
\begin{enumerate}
\item $\mathcal M_t(C)\leq n-\mathcal M_r(C^\perp)$; 
\item for every $\Delta>0$, $\mathcal M_{t+\Delta}(C)\neq n-\mathcal M_r(C^\perp)+1$.
\end{enumerate}
\end{lem}

Before proving it, we first show that this lemma is enough to conclude. Lemma \ref{l2} implies that for every $1\leq r \leq n-k$ and for every $s\geq t$, $$\mathcal M_s(C)\neq n+1-\mathcal M_r(C^\perp).$$ Moreover, for every $s<t$, by the monotonicity property (Theorem \ref{t4}), $$\mathcal M_s(C)<\mathcal M_t(C)< n+1-\mathcal M_r(C^\perp),$$  hence $$\{\mathcal M_s(C) \mid 1\leq s\leq k\}\cap \{n+1-\mathcal M_r(C^\perp) \mid 1\leq r\leq n-k\}=\emptyset.$$ Furthermore, the cardinality of the union $$\{\mathcal M_s(C) \mid 1\leq s\leq k\}\cup \{n+1-\mathcal M_r(C^\perp) \mid 1\leq r\leq n-k\}$$ is equal to $k+n-k=n$ (thanks to the monotonicity property (Theorem \ref{t4}) again).  Since now both sets are included in $\{1,...,n\}$, then $$\{\mathcal M_s(C) \mid 1\leq s\leq k\}\sqcup \{n+1-\mathcal M_r(C^\perp) \mid 1\leq r\leq n-k\}=\{1,...,n\},$$ which completes the proof of Theorem \ref{t1}.\\\\

Let us now prove Lemma \ref{l2} :
\begin{dem}
Let $1\leq r\leq n-k$. 
\begin{enumerate} 
\item We set $t=k+r-\mathcal M_r(C^\perp)$. We want to show that $\mathcal M_t(C)\leq n-\mathcal M_r(C^\perp)$. Let $V\in \Gamma(\mathbb F_{q^m}^n)$ such that $\dim (V\cap C^\perp)\geq r$ and $\dim V=\mathcal M_r(C^\perp)$. We have $$\begin{aligned} \dim(V\cap C^\perp) &= \dim V+\dim C^\perp -\dim(V+C^\perp)\\ & = \mathcal M_r(C^\perp) +n-k -\dim\left(\left(V^\perp \cap (C^\perp)^\perp\right)^\perp\right)\\ & = \mathcal M_r(C^\perp) +n-k -n+ \dim (V^\perp \cap C)\\ &= \mathcal M_r(C^\perp)-k + \dim (V^\perp \cap C).\end{aligned}$$ Since $\dim(V\cap C^\perp)\geq r$, we get that $$ t=r + k - \mathcal M_r(C^\perp) \leq \dim (V^\perp \cap C).$$ Therefore, $$n-\mathcal M_r(C^\perp)=n-\dim V = \dim (V^\perp)\geq \mathcal M_t(C)$$ (since $V\in \Gamma(\mathbb F_{q^m}^n)$, then $V^\perp \in \Gamma(\mathbb F_{q^m}^n)$ by Lemma \ref{l5}).\\

\item We make a proof by contradiction in assuming that there exists some $\Delta>0$, such that $$\mathcal M_{t+\Delta}(C)=n+1-\mathcal M_r(C^\perp).$$ Then there exists $V\in \Gamma(\mathbb F_{q^m}^n)$ such that $\dim(V\cap C)\geq t+\Delta$ and $\dim V=n+1-\mathcal M_r(C^\perp)$. We have $$\dim(V\cap C)=\dim V +\dim C-\dim(V+C)= n+1-\mathcal M_r(C^\perp)+k-(n-\dim(V^\perp\cap C^\perp)).$$ Since $\dim(V\cap C)>t$, we get that $$\begin{aligned} t &< 1-\mathcal M_r(C^\perp)+k + \dim(V^\perp\cap C^\perp)\\ k+r- \mathcal M_r(C^\perp) & < 1-\mathcal M_r(C^\perp)+k + \dim(V^\perp\cap C^\perp) \\ r-1 &< \dim(V^\perp\cap C^\perp). \end{aligned}$$ Since $V^\perp\in \Gamma(\mathbb F_{q^m}^n)$ by Lemma \ref{l5} and  $\dim(V^\perp\cap C^\perp)\geq r$, we have $$\dim V^\perp \geq \mathcal M_r(C^\perp).$$ 

However, $$\dim V^\perp = n-\dim V= n-(n+1-\mathcal M_r(C^\perp))=\mathcal M_r(C^\perp)-1,$$ which contradicts the previous inequality.\\
\end{enumerate}
This completes the proof of Lemma \ref{l2} and that of Theorem \ref{t1}. $\square$\\
\end{dem}
\end{dem}

We can then derive from Theorem \ref{t1} the following characterization of the $r$-MRD codes in terms of the rank weight of the dual code :
\begin{cor}
Keeping notation as in Theorem \ref{t1}, for every $1\leq r \leq k$, the code $C$ is $r$-MRD if and only if $\mathcal M_r(C^\perp)\geq k-r+2$.
\end{cor}

\begin{dem}
Let $1\leq r\leq k$. Assume first that $\mathcal M_r(C)=n-k+r$. By monotonicity property (Theorem \ref{t4}), for all $r\leq s\leq k$, we have $\mathcal M_s(C)=n-k+s$. Hence, for all $r\leq s\leq k$, $$n+1-\mathcal M_s(C)=n+1-(n-k+s)=k-s+1$$ and by Theorem \ref{t1}, $$\{1,2,\ldots,k-r+1\}\subseteq \{1,...,n\}\setminus \{\mathcal M_t(C^\perp)\mid 1\leq t\leq n-k\}.$$ It implies that $d(\lambda(C^\perp))=\mathcal M_1(C^\perp)\geq k-r+2$.  \\

Conversely, assume that $d(\lambda(C^\perp))\geq k-r+2$. By monotonicity property (Theorem \ref{t4}), it means that $$\{1,...,k-r+1\}\cap \{\mathcal M_t(C^\perp)| 1\leq t\leq n-k\}=\emptyset$$ and Theorem \ref{t1} implies that $$\{1,...,k-r+1\}\subseteq \{n+1-\mathcal M_s(C)| 1\leq s\leq k\}.$$ Finally, again by the monotonicity property (Theorem \ref{t4}), we obtain that \begin{center} $\mathcal M_k(C)=n$, $\mathcal M_{k-1}(C)=n-1$,$\ldots$,  $\mathcal M_r(C)=n+1-(k-r+1)=n-k+r$\end{center} which proves that $C$ is $r$-MRD. $\square$\\
\end{dem}

\section*{Acknowledgments}
This research was supported by the Singapore National Research Foundation under Research Grant NRF-RF2009-07. The author would like to warmly thank Prof. F. Oggier for introducing him to this nice topic and for her meaningful advice and careful read-through of the paper.

\end{document}